\documentclass[prd,draft, floats,noshowpacs,nofootinbib]{revtex4}

\def\half{{\textstyle{\frac{1}{2}}}}

\def\eq#1{eq.~(\ref{#1})}

\def\eqs#1#2{eqs.~(\ref{#1})\ and (\ref{#2})}

\def\eg{\hbox{\it e.g.}}

\def\gesim{\,{\raise-3pt\hbox{$\sim$}}\!\!\!\!\!{\raise2pt\hbox{$>$}}\,}

\def\lesim{\,{\raise-3pt\hbox{$\sim$}}\!\!\!\!\!{\raise2pt\hbox{$<$}}\,}

\def\boldoverdot{\,{\raise6pt\hbox{\bf.}\!\!\!\!\>}}

\def\lcal{{\cal L}}

\def\diag{\hbox{\diag}}

\def\doubleundertext#1{

{\undertext{\vphantom{y}#1}}\par\nobreak\vskip-\the\baselineskip\vskip4pt%

\undertext{\hbox to 2in{}}}

\def\inbox#1{\vbox{\hrule\hbox{\vrule\kern5pt

     \vbox{\kern5pt#1\kern5pt}\kern5pt\vrule}\hrule}}

\def\sqr#1#2{{\vcenter{\hrule height.#2pt

      \hbox{\vrule width.#2pt height#1pt \kern#1pt

         \vrule width.#2pt}

      \hrule height.#2pt}}}

\def\today{\ifcase\month\or
  January\or February\or March\or April\or May\or June\or
  July\or August\or September\or October\or November\or December\fi
  \space\number\day, \number\year}

\def\pmb#1{\setbox0=\hbox{#1}%

  \kern-.025em\copy0\kern-\wd0

  \kern.05em\copy0\kern-\wd0

  \kern-.025em\raise.0433em\box0 }

\def\sumprime_#1{\setbox0=\hbox{$\scriptstyle{#1}$}

  \setbox2=\hbox{$\displaystyle{\sum}$}

  \setbox4=\hbox{${}'\mathsurround=0pt$}

  \dimen0=.5\wd0 \advance\dimen0 by-.5\wd2

  \ifdim\dimen0>0pt

  \ifdim\dimen0>\wd4 \kern\wd4 \else\kern\dimen0\fi\fi

\mathop{{\sum}'}_{\kern-\wd4 #1}}

\newcommand{\nc}{\newcommand}
\nc{\beq}{\begin{equation}}  \nc{\eeq}{\end{equation}}
\nc{\bea}{\begin{eqnarray}}  \nc{\eea}{\end{eqnarray}}
\nc{\baa}{\begin{array}}     \nc{\eaa}{\end{array}}
\nc{\bit}{\begin{itemize}}   \nc{\eit}{\end{itemize}}
\nc{\ben}{\begin{enumerate}} \nc{\een}{\end{enumerate}}
\nc{\bce}{\begin{center}}    \nc{\ece}{\end{center}}

\begin{document}
\rightline{MCTP-02-71}

\title{Instantons and SL(2,R) Symmetry in Type IIB Supergravity}
\author{\sc Martin B. Einhorn}
\affiliation{Michigan Center for Theoretical Physics,\\ Randall
Laboratory, The University of Michigan,
Ann Arbor, MI 48109-1120\\E-mail: {\tt meinhorn@umich.edu}}
\date{\today}

\begin{abstract}
We discuss the relation between the dual formulations of Type IIB
supergravity emphasizing the differences between Lorentz and Euclidean
signature.  We demonstrate how the SL(2,R) symmetry of the usual action
is manifested in the solution of the equations of motion  
with Euclidean signature for the dual theory.
\end{abstract}
\maketitle

\section{Instantons and SL(2,R) Symmetry in Type IIB SUGRA}

It is well-known that the Lagrangian for Type IIB SUGRA possesses an
SL(2,R) invariance.\cite{GSW2,Polchinski2}  This symmetry is however only
evident as a local symmetry when the RR potentials are chosen to be counterparts of the NS
fields, the dilation $\phi$ and the antisymmetric tensor $B_{\mu \nu}$.  By this, we mean that the RR potentials must be chosen
to be the scalar $a$ and the two-form $(C_{\mu \nu})$.  
If we denote an element of SL(2,R) as 
\begin{eqnarray}
\left(\begin{array}{cc}
p & q \\
r & s
\end{array}
\right),
\end{eqnarray}
then $(B_{\mu
\nu}, C_{\mu \nu})$ transform linearly as an SL(2,R) doublet,
while $\tau=a+i\exp(-\phi)$ transforms non-linearly via the modular transformation 
\begin{eqnarray}\label{modular}
\tau \rightarrow \frac{p\tau+q}{r\tau+s}.
\end{eqnarray}

Type IIB SUGRA possesses a $D= -1$ brane, an instanton.\cite{GGP,
einhorn1}  This is unique among D-branes in several respects --- all
other D-branes are static solutions of the classical fields equations,\cite{DuffKhuriLu}
so they take precisely the same forms in a space-time with Lorentzian
(or pseudo-Riemannian) signature or, after Wick rotation, in a
space-time with Euclidean (or Riemannian) signature.  In contrast, the
instanton is a finite action solution of the Euclidean field equations 
that depends non-trivially on all coordinates, including the Euclidean
time.  Although interpreted as a tunneling amplitude between states
in the Hilbert space that is built up in Lorentzian space-time, it only exists
as a solution of the field equations with Euclidean signature.

Under these circumstances, it is not so clear that the duality
characteristic of other D-branes applies to the $D= -1$ brane or, if it
does apply, how it is to be formulated.  In particular, how the SL(2,R) symmetry 
can be seen in the dual, magnetic picture is not at all apparent.
In a previous paper discussing the instanton solution,\cite{einhorn1} 
we have discussed some of the relations between Lorentz signature (LS) 
and Euclidean signature (ES).  In  this note, we will concentrate on the role 
of the SL(2,R) symmetry.  

As mentioned at the outset, the SL(2,R) symmetry is manifested in the
formulation of Type IIB SUGRA in which the RR potentials have the same
tensor structure as their NS counterparts.  Since we are exclusively 
interested in the instanton, we will suppress the two-form fields for
simplicity.  The Lagrangian density in the Einstein frame is then simply
\begin{equation}\label{axionlagrang}
{\cal L}_0 = -R +\half{(\nabla\phi)}^2 +\half{e^{2\phi}}{(\nabla a)}^2
\end{equation}
 The last two terms, the ``matter" Lagrangian, may also be written as 
\begin{equation}\label{taulagrang}
\lcal_{0m}\equiv \half\frac{ \nabla^\mu \tau \nabla_\mu \tau^*}{{\tau_2}^2},
\end{equation}
 where $\tau\equiv \tau_1 + i\tau_2\equiv a + i e^{-\phi}.$
 From the point of view of the instanton, this is the
``electric'' picture, in which the instanton would be seen as an
``elementary'' source coupled to the RR field $a.$\footnote{The notion 
of an ``elementary" instanton is rather radical, but in this case, the source 
is to be interpreted as a boundary condition.\cite{einhorn1}} 

To realize the
instanton in its usual form as an extended solution of the 
source-free field equations requires the dual, or ``magnetic,"
formulation in which the dynamical variable $a$ is exchanged for an
eight-form potential $C_8$, with Lagrangian 
\begin{equation}\label{ceightlagrang}
{\cal L}_8 = -R+\half{(\nabla\phi)}^2 +
{\textstyle \frac{1}{2\cdot9!}} \displaystyle e^{-2\phi} F_9^2
\end{equation}
The nonlocal relationship between the fields $a$ and $C_8$ is implicit in the relation 
between their associated field strengths 
\begin{equation}\label{dual}
F_9 = e^{2\phi}\;{}^*F_1
\end{equation}
 where $F_1\equiv da,$ $F_9\equiv dC_8,$ and the ${}^*$ denotes the usual Hodge dual.  This
relation is metric dependent and {\it assumes Lorentzian signature.}
In form notation, the ``matter'' action above may be expressed as
\begin{equation}\label{axionmatter}
{\cal L}_{0m} = \half  d\phi \wedge {}^* d\phi 
+ \half{e^{2\phi}} da \wedge {}^* da
\end{equation}
 The corresponding dual action is 
\begin{equation}\label{c8matter}
{\cal L}_{8m} =\half  d\phi\wedge {}^* d \phi +
\half  {e^{-2\phi}} d C_8\wedge {}^* dC_8
\end{equation}
 The formal substitution in \eq{dual}, $dC_8 \rightarrow e^{2\phi}{}^* da$ 
does \underline{not} map the two actions into each other,
owing to the fact that $^{* *}F_p ={(-)}^{p+s} F_p$, where $s$
is the signature of the metric.\footnote{Here, we have assumed even
space-time dimension since we are primarily interested in $D=\:10$.}  As a result 
\begin{equation}\label{duality}
e^{-2\phi}F_9 \wedge {}^* F_9 ={(-)}^s e^{2\phi} F_1 \wedge {}^* F_1.
\end{equation}
so that, for Lorentzian signature, the substitution results in a wrong sign kinetic energy.
However, the equations of motion (EOM) determined from 
these two actions \underline{are} mapped into each other (again, only for Lorentzian signature).  
Duality is a property of the EOM, not of the action.

To be precise, the EOM stemming from \eq{axionmatter} are 
\begin{eqnarray}\label{axionEOM}
\nabla_\mu(e^{2\phi}\nabla^\mu a) &=& 0\nonumber\\
\nabla^2\phi - e^{2\phi}{(\nabla a)}^2 &=&0
\end{eqnarray}
The EOM in the $C_8$ formulation, \eq{c8matter} are
\begin{eqnarray}\label{ceightEOM}
\nabla_\mu (e^{-2\phi} F^{\mu\mu_1...\mu_8}{\big )}  &=& 0\nonumber \\
\nabla^2 \phi + e^{-2\phi} \frac{{F_9}^2}{\textstyle 9!}  &=& 0
\end{eqnarray}
 These EOM are formally valid for either LS or ES.  For LS, under
the substitution \eq{dual}, the latter equations become the former.\footnote{Because 
the metric in the Einstein frame is not transformed by SL(2,R),
we have suppressed Einstein's equations here.  The instanton solution of course
satisfies all the EOM.\cite{einhorn1}}

If, on the other hand, we consider these same  
two Lagrangians for Euclidean signature, then they \underline{are} 
mapped into one another under the formal substitution \eq{dual}, 
but the EOM are \underline{not}.  
To see this explicitly, consider the solution of \eq{c8matter}.  
In the absence of any seven-branes, 
there are no source terms, so the first equation implies
\begin{equation}\label{axionb}
e^{-2\phi}F_9 =db
\end{equation}
 for some scalar $b$, and the second equation then becomes
\begin{equation}\label{bEOM1}
\nabla^2 \phi +e^{2\phi}{(\nabla b)}^2 =0.
\end{equation}
 The Bianci identify $dF_9 =0$ then implies
\begin{equation}\label{bEOM2}
\nabla_\mu(e^{2\phi}\nabla^\mu b) =0.
\end{equation}
If we compare \eqs{bEOM1}{bEOM2} with \eq{axionEOM}, we see that the EOM for ES for the scalar $b$ is the same as if we replaced $a$ by $ib$.  For this reason, it is sometimes stated that the pseudoscalar 
$a\rightarrow ia$ under a Wick rotation.\cite{pvn}  This mnemonic is misleading, 
because this rule may only be used to obtain the form of the Euclidean EOM and \underline{not} for the value of the
action nor for variations of the action other than first order (\eg, to examine whether the solution is a local minimum in field space.)
 These EOM for ES are \underline{as if} the action were
\begin{equation}\label{pseudoaction}
S=\half \int d^{10}x\sqrt{g}\left[ (\nabla\phi)^2 - e^{2\phi}(\nabla b)^2\right]
\end{equation}
 However, the \underline{value} of the matter action is correctly given by 
\begin{equation}\label{svalue}
S_m =\half  \int d^{10}x\sqrt{g} \left[ (\nabla\phi)^2+e^{2\phi} (\nabla b)^2\right] \ge 0.
\end{equation}
Nor are variational derivatives beyond the first correctly generated by \eq{pseudoaction}.
 For these reasons, we will refer to \eq{pseudoaction} as the pseudo-action.

Given the clash between the EOM for Euclidean Signature in the dual
formulations, it is not so clear that the instanton can be represented
as an ``elementary'' excitation for \eqs{axionlagrang}{taulagrang}.  In particular, the
SL(2,R) symmetry of Eq.~(2) is not obviously a Noether symmetry of 
the pseudoaction \eq{pseudoaction}.  Of course, any given instanton,
like any other D-brane, will break the symmetry owing to its collective coordinates such as its origin.
Whether different instantons solutions are related by an SL(2,R) 
transformation is not at all clear.  

The symmetry of the solutions of
the magnetic EOM's are the same as the symmetries of the pseudo-action\eq{pseudoaction}.
Even though this is not the true action, it suffices to examine
whether this action possesses an SL(2,R) symmetry.  Moreover, since
conserved currents only depend on the EOM, the symmetries of the
pseudo-action may be used to determine them via the usual Noether procedure.
In order to see that the pseudoaction \underline{does} in fact possess an SL(2,R) symmetry, 
let us observe the following little lemma:  The expression
\begin{equation}
-\frac{2\nabla_\mu W \nabla^\mu Z}{{(W-Z)}^2}
\end{equation}
 is SL(2,R) invariant if $W$ and $Z$ are any two (real or
complex) scalar fields that transform by the modular transformation \eq{modular}.
It can easily be shown by explicit calculation that $\nabla
W \rightarrow\frac{\nabla W}{{(r W + s)}^2}$ and similarly for $Z$,
where $(W - Z)\rightarrow \frac{W - Z}{(r W + s)(r Z + s)}$.
In the case of the Lagrangian \eq{axionlagrang}, SL(2,R) is realized for $W = \tau$,
$Z = \tau^*$.\footnote{Because $p,q,r,s$ are real, $\tau$ and $\tau^*$ transform the same way.}
For the pseudo-action \eq{pseudoaction}, SL(2,R) is realized by
taking $W =i(b + e^{-\phi}),$ $ Z = i(b-e^{-\phi})$. (Requiring
$W \& Z$ to transform by \eq{modular}, one may deduce the transformation
properties of $b$ and $e^{-\phi}$.)  In other
words, SL(2,R) symmetry \underline{is} in a sense preserved is by the replacement of $a \rightarrow
i b$, provided one makes the replacement in $\tau$ and $\tau^*$ independently.
This may be seen as an automorphism of the original SL(2,R) which may be
obtained by the formal replacement $q\rightarrow i q, r\rightarrow -i r,$ with $ps - qr =1$ as before.

For completeness, let us record the three SL(2,R) currents for each
form.  In the original form \eq{axionlagrang}, we may choose these as
\begin{eqnarray}
J^\mu &=&e^{2\phi}\nabla^\mu a \\
K^\mu &=& \nabla^\mu \phi - a J^\mu \\
L^\mu &=& (a^2 +e^{-2\phi})J^\mu + 2a K^\mu
\end{eqnarray}
 The motivation for the particular choices here is that 
$J^\mu$ is associated with the shift
$a\rightarrow a + c,$ $K^\mu$ with the scaling $e^\phi
\rightarrow e^\nu e^\phi, a \rightarrow e^{-\nu}a$.   
The third is inherently non-linear.  We may 
express these in a more familiar language, using the isomorphism of
the SL(2,R) algebra with the three-dimensional Lorentz group SO(2,1):
\begin{equation}
[J, K_1]= i K_2,\;\;[J, K_2]= -i K_1,\;\;[K_1, K_2]= -i J.
\end{equation}
 Then the generators associated with each current transform
as follows:
\begin{eqnarray}
J^\mu &\sim&\;\;\half(K_2 + J) \\
K^\mu &\sim&\;\;K_1 \\
L^\mu &\sim&\;\;\half(K_2 - J)
\end{eqnarray}
 Correspondingly, for the \underline{solution} of the dual
theory, we have the conserved currents 
\begin{eqnarray}
J^\mu &=& e^{2\phi}\nabla^\mu b \\
K^\mu &=& \nabla^\mu\phi + bJ^\mu \\
L^\mu &=& (e^{-2\phi}- b^2)J^\mu + 2bK^\mu.
\end{eqnarray}
 These of course are only conserved for the source-free EOM
in either case.

All the instanton solutions obtained in Ref.~\cite{einhorn1} are related to each other by $SL(2,R).$ Thus, any change in the three integration constants discussed there may be interpreted as an SL(2,R) transformation.  These relations and currents are useful for understanding the role of the instantons and the properties of the ground state.   However, inasmuch as variations higher than the first are not correctly obtained from the pseudoaction, it remains for future work to determine the implications of this symmetry for transition amplitudes mediated by instantons.  

\section{Acknowledgments}
I would like to thank A. Mecke for discussions.

\end{document}